\documentclass[prl,floats,twocolumn,showpacs,floatfix,superscriptaddress]{revtex4}
\usepackage{graphics,graphicx,dcolumn,bm,fleqn,epic,eepic,float}
\usepackage{amssymb,amsmath,multirow,rotate,color,float,mciteplus}
\usepackage[latin1]{inputenc}

\usepackage[dvips]{epsfig}

\begin{document}

\title{Electro-hydrodynamics near Hydrophobic Surfaces }

\author{S. R. Maduar}
\affiliation{A.N.~Frumkin Institute of Physical
Chemistry and Electrochemistry, Russian Academy of Sciences, 31
Leninsky Prospect, 119071 Moscow, Russia}
\affiliation{Department of Physics, M. V. Lomonosov Moscow State University, 119991 Moscow, Russia}

\author{A. V. Belyaev}

\affiliation{A.N.~Frumkin Institute of Physical
Chemistry and Electrochemistry, Russian Academy of Sciences, 31
Leninsky Prospect, 119071 Moscow, Russia}

\affiliation{Department of Physics, M. V. Lomonosov Moscow State University, 119991 Moscow, Russia}

\author{V. Lobaskin}
\affiliation{School of Physics, University College Dublin, Belfield, Dublin 4, Ireland}

\author{O. I. Vinogradova}
\email[Corresponding author: ]{oivinograd@yahoo.com}
\affiliation{A.N.~Frumkin Institute of Physical
Chemistry and Electrochemistry, Russian Academy of Sciences, 31
Leninsky Prospect, 119071 Moscow, Russia}
\affiliation{Department of Physics, M. V. Lomonosov Moscow State University, 119991 Moscow, Russia}
\affiliation{DWI - Leibniz Institute for Interactive Materials, RWTH Aachen, Forckenbeckstra\ss e 50, 52056 Aachen,
  Germany }

\newcommand\Xsin{\mbox{sin}}
\newcommand\Xcos{\mbox{cos}}
\newcommand\Xsec{\mbox{sec}}
\newcommand\Xlog{\mbox{ln}}

\newcommand*{\mycommand}[1]{\texttt{\emph{#1}}}
\def\p{\par $\bullet$ }

\date{\today}

\begin{abstract}

We show that an electro-osmotic flow near the slippery hydrophobic surface
 depends strongly on the mobility of  surface charges, which are balanced by counter-ions of the electrostatic diffuse layer. For a hydrophobic surface with immobile
charges the fluid transport is considerably amplified by the
existence of a hydrodynamic slippage. In contrast, near the hydrophobic surface
with mobile adsorbed charges it is also controlled by an additional electric force, which increases the shear
stress at the slipping interface. To account for this we formulate electro-hydrodynamic boundary conditions at the slipping interface, which should be applied to quantify electro-osmotic flows instead of hydrodynamic boundary conditions. Our theoretical predictions are fully supported by dissipative particle
dynamics simulations with explicit charges. These results lead to a new interpretation of zeta-potential of hydrophobic surfaces.

\end{abstract}
\pacs {47.57.jd, 83.50.Lh, 68.08.-p}


\maketitle

 The electrostatic diffuse layer (EDL), i.e. the region where the surface
charge~\cite{note2} is balanced by the cloud of counter-ions, is the central concept in understanding dynamic properties of colloidal systems since it is an origin of numerous electrokinetic effects. This includes electro-osmotic (EO) flow with respect to a charged surface that provide an extremely efficient way to drive and manipulate flows in micro- and nanofluidic devices~\cite{squires2005,schoch.pb:2008,eijkel.jct:2005}. Most studies of EO assume no-slip hydrodynamic boundary conditions at the surface, which are typical for wettable (hydrophilic) surfaces. In this situation the \emph{outer} EO velocity $u_1$ (outside of the \emph{thin} EDL) due to the tangential electric field $E_t$ is given by the  Smoluchowski formula
\begin{equation}\label{smoluchovsky}
    {u_1}=-\frac{E_t q_1}{\eta \kappa},
\end{equation}
where $\eta$ is the viscosity of the solution, $ q_1$ is the charge density at the  no-slip surface,
 related to the so-called zeta-potential, $\zeta_1=q_1/\kappa\varepsilon$.
Here, $\varepsilon$ is the permittivity of the solution, and
$\kappa=\lambda_D^{-1}$ is the inverse Debye screening length. Obviously, $\zeta_1$ is equal to the EDL potential.

In practice, however, non-wetting (hydrophobic) materials show hydrodynamic slip, characterized by the slip length $b$ (the distance within the solid at which the flow profile extrapolates to zero)~\citep{vinogradova1999}. Some moderate slip, of the order of nm, was detected even in weakly hydrophilic systems~\citep{bocquet.l:2010}.  For a charge density $q_{2}$  the slipping solid interface, simple arguments show that the outer EO velocity is given by ~\cite{muller.vm:1986,joly2004}:
\begin{equation}\label{isotropic}
      u_{2}= - \frac{E_t q_{2}}{\eta \kappa} (1 + b \kappa)
\end{equation}
The zeta potential  was thus defined as  $\zeta_{2}=q_{2} (1+b\kappa)/\kappa\varepsilon$. Since at hydrophobic solids $b$ can be of the order of tens of nanometers ~\citep{vinogradova:03,charlaix.e:2005,joly.l:2006,vinogradova.oi:2009}, for typically nanometric Debye length some small enhancement of the zeta potential and EO flow was observed experimentally~\cite{bouzigues.c:2008}. We remark however that Eq.(\ref{isotropic}) fails to predict a realistic $\zeta_{2}$ of the free interface of bubbles ($b=\infty$) or oil drops~\cite{creux.p:2009,marinova.kg:1996,Takahashi05} and, in fact, of systems with large partial slip such as gas sectors of superhydrophobic surfaces~\cite{nizkaya.tv:2014}.
\begin{figure}
  \begin{center}
  \includegraphics[width=7 cm,clip]{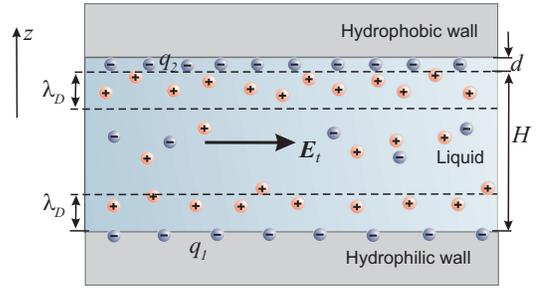}\\
  \end{center}
   \vspace{-0.4cm}
  \caption{Sketch of the system in the case of an asymmetric channel with one no-slip wall. In the case of a symmetric channel both slippery walls are equal. Fixed surface charges are in the walls, mobile surface charges are adsorbed to the neutral walls. This is important for a definition of $H$ in simulations, but not in the theory, where ions are point-like, so that $d$ is infinitesimally thin.  }\label{fig:geometry}
\end{figure}

Previous studies assumed that electric charge associated with slippery surface was \emph{immobile}, which is not justified for bubbles or drops. This is also by no means obvious for hydrophobic solids, as the `gas cushion' model of apparent hydrophobic slip relates it to the formation of a depletion layer at the surface~\cite{vinogradova.oi:1995a}.
This idea  has received a microscopic foundation in terms of a prewetting transition~\cite{andrienko.d:2003},
and was confirmed by recent simulations~\cite{dammer.sm:2006,sendner.c:2009}.
There is a growing evidence that such an interface is weakly charged~\cite{kirby2008,pushkarova.ra:2008}. The existence of surface charges can be caused by ion adsorption~\cite{Tobias.d:2013,Bocquet_ionspecific}, so that they are laterally \emph{mobile}, and can respond to the external electric field. Indeed, experiments with foam films suggested that electric field indeed drives charges at the free surface in opposite directions~\cite{bonhomme.o:2013}, and recent analysis has shown that this could enhance the shear stress~\cite{choi.w:2010}, but we are unaware of any prior work that has explored what happens in channels with partially slipping hydrophobic walls if adsorbed charges are mobile.
In this Letter, we present some general theoretical arguments and results of Dissipative Particles Dynamics (DPD) simulations, which allow us to quantify an EO flow in such channels. Our analysis leads to a new interpretation of zeta-potential of hydrophobic surfaces.

We consider an EO flow of an aqueous electrolyte solution between two flat walls as sketched in Fig.~\ref{fig:geometry}, and define the locus of surfaces at beginning of  EDLs, $z=0$ and $z=H \gg \lambda_D$, where EO slip velocities and zeta-potentials are determined. The hydrophilic surface has a density of charge $q_1$, and the hydrophobic surface is characterized by the density of charge $q_2$. 

We keep our theory at the mean-field level and treat ions as point-like. Let us first consider an asymmetric channel with one non-slipping hydrophilic surface ($z=0: u=0$).  To describe the fluid velocity at the hydrophobic wall we suggest a boundary condition, which takes into account that the tangential stress balance represents a combination of both hydrodynamic and Maxwell stress components~\cite{epaps}
\begin{equation}  \label{BCv_sliplength}
  z=H: \quad u = b (-\partial_z u + (1-\mu) q_2 E_t /\eta),
\end{equation}
where parameter $\mu$ can vary from $0$ for fully mobile charges to $1$ in the case of fixed charges. Now we want to compute the velocity profile, which would be expected within a continuous theory when condition (\ref{BCv_sliplength}) is valid.

The fluid flow satisfies Stokes' equations with an electrostatic body force
\begin{equation}\label{Stokes}
    \eta\nabla^2 \emph{\textbf{u}} = \varepsilon\nabla^2 \psi \textbf{\emph{E}}, \quad \nabla \cdot\textbf{\emph{u}}=0,
\end{equation}
where electric field represents a superposition of an external and a created by surface charges fields
$\emph{\textbf{E}} =\textbf{\emph{E}}_t-\nabla\psi$.
The solution of Eq.(\ref{Stokes}) together with the Poisson-Boltzmann equation and prescribed boundary conditions in general requires a numerical method. However, in case of typical for hydrophobic surfaces low surface potentials $\psi(z)$ can be obtained analytically~\cite{epaps}. In the thin EDL limit we then predict an outer EO `shear' flow:

\begin{equation}\label{shear}
   \frac{u(z)}{u_1} = 1 + \frac{z}{b+H} \left[(1+ \mu \kappa b )q_2/q_1  -1 \right]
   \end{equation}
The apparent EO slip at the hydrophobic surface is then
\begin{equation}\label{zeta_general}
   \frac{u_2}{u_1}  =  1 - \frac{1 - (1+ \mu b \kappa )q_2/q_1 }{1+b/H},
\end{equation}
which suggests immediately that it is not \textbf{its} unique characteristic. In contrast, it depends strongly on the second surface of the channel provided $b$ is of the order of $H$ or larger. One striking prediction is that even uncharged hydrophobic surface, $q_2=0$, can
induce an apparent EO slip. Another important result is that  Eq.(\ref{zeta_general}) even at $\mu=1$ differs from Eq.(\ref{isotropic}) derived for a single interface and suggests that at $b/H \gg 1$ the EO slip velocity  becomes independent on $b$ and saturates to $u_2/u_1=1+\kappa H q_2/q_1$. However, when $\mu=0$ this large slip limit inevitably leads to $u_2/u_1=1$.

Now, the same strategy can be  applied for a symmetric hydrophobic channel (with the charge density $q_2$ and slip length $b$ at both walls), which is also relevant for free soap and foam films that are currently a subject of active research~\cite{bonhomme.o:2013,joly.l:2014}. We apply a symmetry condition ($z=H/2: \partial_z u =0$) together with Eq.(\ref{BCv_sliplength}) to solve Eq.(\ref{Stokes}) in the thin EDL limit, and conclude that two situations occur.
For a finite slip we obtain~\cite{epaps}
\begin{equation}\label{symmetric}
      u_{2}= -\frac{E_t q_2}{\eta \kappa} (1 + \mu b \kappa)
\end{equation}
Eq.(\ref{symmetric}) reduces to Eq.(\ref{isotropic}) when $\mu=1$ and justifies the use of the Smoluchowsky equation when $\mu=0$. For $b=\infty$ and $\mu=0$ we predict zero flow rate in the channel with a vanishing at very large $\kappa H$ outer EO velocity~\cite{epaps},
 \begin{equation}\label{infinite}
      u_{2}= -\frac{E_t q_2}{\eta \kappa}\frac{2}{\kappa H}
\end{equation}
which explains simulation data for this case~\cite{huang.dm:2008b}.

 In order to assess the validity of the above approach we employ DPD simulations~\cite{HOOGERBRUGGE.P:1992,Espanol1995,Groot1997} carried out using the open source package ESPResSo~\cite{ESPResSo} (details are given in ~\cite{epaps}). We use a simulation cell confined between two impermeable walls always located at $z=0$ (except the case of a symmetric hydrophobic channel with mobile surface charges, where the lower wall was at $z=-1$) and $14 \sigma$, where $\sigma$ sets the length scale. The value of $\kappa = (8 \pi \ell_B c_0)^{1/2}$ with Bjerrum length $\ell_B=e^2/4 \pi \varepsilon  k_BT$  was determined by using the concentration, $c_0\simeq 5\times10^{-2}\sigma^{-3}$, outside EDLs, which  gives $\kappa^{-1}=1-1.2 \sigma$ and provides large $\kappa H$. We set-up $b$ from 0 to $\infty$ at the walls by using a tunable slip method~\cite{Smiatek2008,epaps}.

Immobile surface charges are implemented by randomly placing discrete unit charges $q_se$ in the no-slip hydrophilic walls, to provide
homogeneous $q_1 = 0.15 q_se/\sigma^{2}$. We adjusted $4\pi \ell_B q_1 / \kappa< 1$ to ensure the `weak charge' behavior. Fixed charges of a density $q_2$ at the hydrophobic wall are created similarly.

\begin{figure}
\begin{center}
\includegraphics[width=4cm]{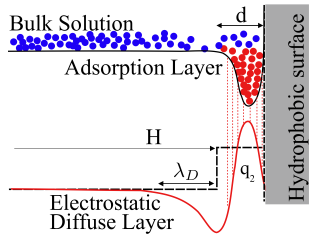}
\end{center}
\caption{Top: LJ adsorption potential applied in simulations. Bottom:  A concentration profile of adsorbed ions and the model with homogeneous charge distribution inside the adsorbed layer. }\label{finite_thickness}
\end{figure}
 The mobile charges are modeled by applying  an effective interfacial potential, which leads to selective adsorption of one type of ions to an  electro-neutral hydrophobic wall. Namely, we set Lennard-Jones (LJ) potential between negative ions and the hydrophobic wall (see Fig~\ref{finite_thickness}), since it qualitatively reproduces the potentials of mean force for surface active ions~\cite{Netz.r:2012}.  The density of adsorbed charge, $q_2$, can be regulated by the strength of LJ potential. The ratio $q_2/q_1$ was varied from  $1\pm0.03$ to $3\pm0.1$ by setting different values of $q_1$ at the no-slip (hydrophilic) surface. Fixed in such a way, adsorbed charges are confined in a  layer of a thickness $d$ being in thermodynamic equilibrium with the bulk electrolyte solution and respond to $E_t$. The thickness of the adsorbed layer,  $d\simeq\sigma$, is determined from the simulation data~\cite{epaps}, and the locus of surfaces was at $z\simeq13 \sigma$ or $z=0$.

\begin{figure}
\center{\includegraphics[scale=0.60]{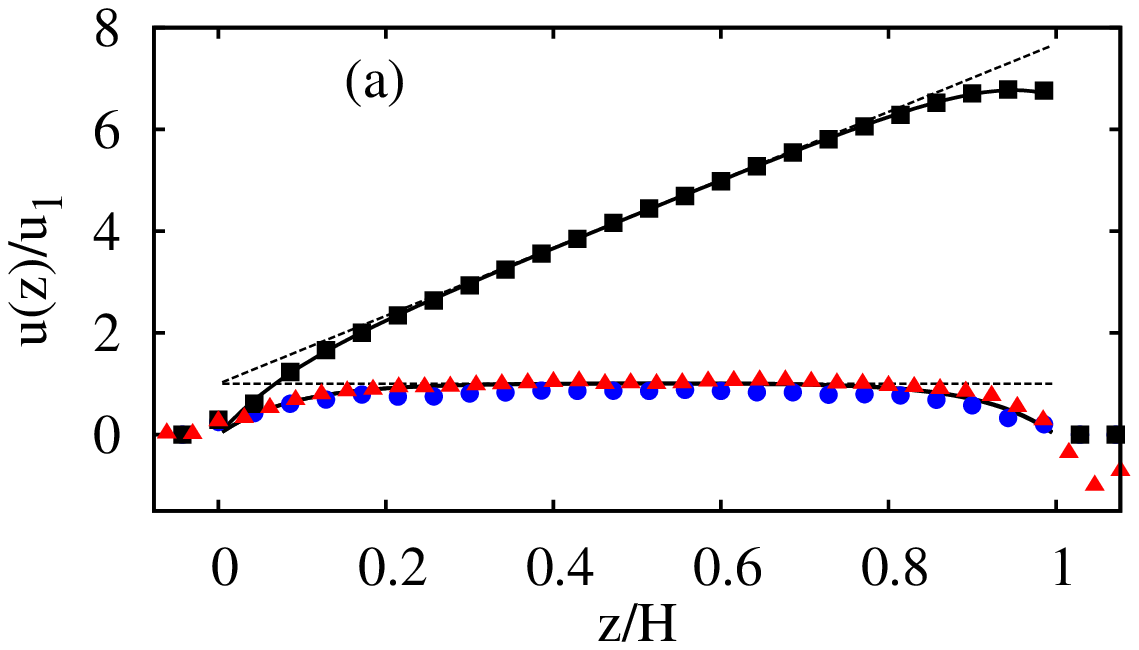}}\\
\center{\includegraphics[scale=0.605]{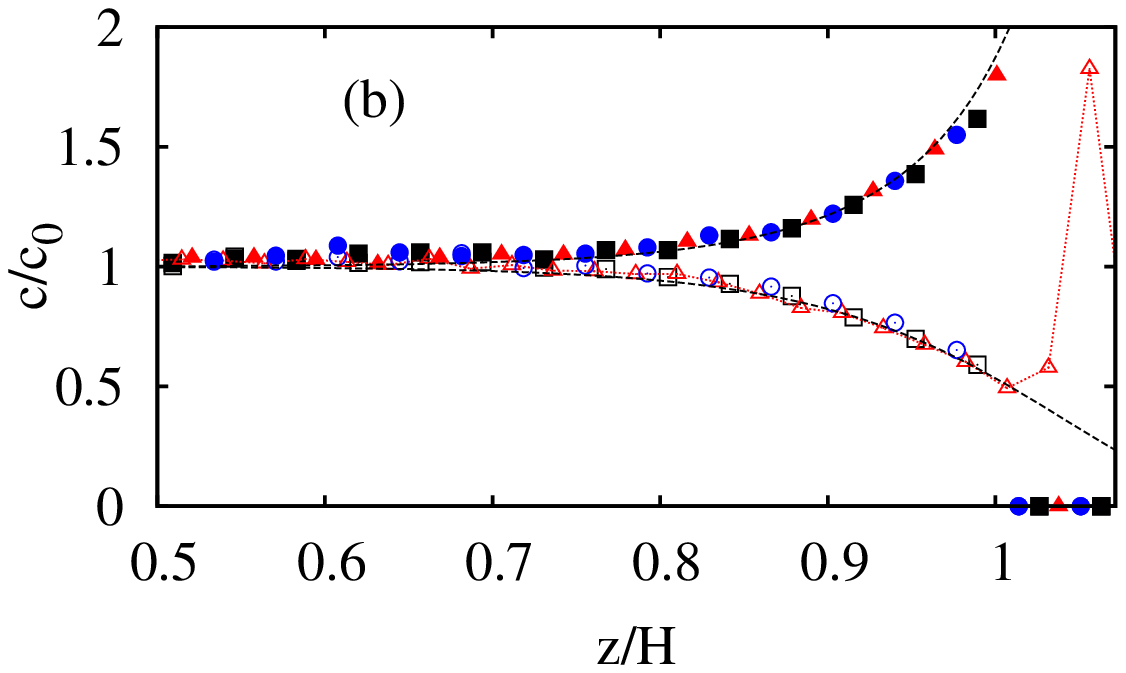}}
\caption{(a) Fluid velocity profiles  simulated at $q_2/q_1 = 1$ and ${\kappa H = 12}$ (symbols). Circles correspond to a hydrophilic channel, squares and triangles - to a channel with a hydrophobic surface, $b/H=1.2$, with $\mu=1$ and 0. Solid curves show solutions of linearized Eq.(\ref{Stokes}), dotted lines - predictions of Eq.(\ref{shear}); (b) Corresponding cation (filled symbols) and anion (open symbols) profiles with the theoretical expectations (dotted curves).}
\label{fig:test}
\end{figure}

We begin by studying an asymmetric channel, where a variety of situations occurs depending on the parameters of the surfaces. Fluid velocity profiles $u(z)$ were first simulated with $q_2/q_1=1$, ${\kappa H = 12}$, and $b/H=1.2$, by setting mobile ($\mu=0$) and immobile ($\mu=1$) charges at the slipping surface. The results are shown in Fig.~\ref{fig:test}(a). Also included are the data obtained for a channel with two hydrophilic walls ($b=0$). A general conclusion from this plot is that the simulation results are in excellent agreement with predictions of mean-field theory, confirming the validity of a continuum description and electro-hydrodynamic boundary condition, Eq.(\ref{BCv_sliplength}). For a hydrophilic channel we observe a classical behavior, where the \emph{inner} fluid velocity in the EDL increases from zero on the surfaces with high gradients to form an outer `plug' EO flow in the electro-neutral center. A hydrophobic slippage strongly amplifies the velocity if surface charges are immobile with an outer `shear' flow, perfectly described by Eq.(\ref{shear}). The slipping surface with mobile charges generates a `plug' profile in the center, and neither outer nor inner EO velocities show a manifestation of the hydrodynamic slip.  Simulation data show that this is however accompanied by some negative `flow' of the  adsorbed layer. We finally note that simulated ion density profiles  are superimposed in all cases as seen in Fig.~\ref{fig:test}(b). This confirms that the EO slip near hydrophobic surfaces no longer reflect the sole EDL potential.

\begin{figure}
\centering
  \includegraphics[scale=0.60]{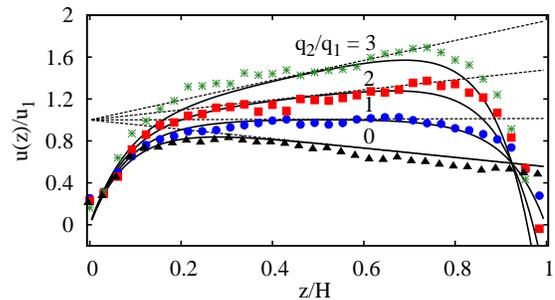}
  \caption{Fluid velocity profile simulated at  $\kappa H = 12$, $b/H = 1.2$, $q_2/q_1 = 0, 1, 2, 3$, and $\mu=0$.  Solid curves show  theoretical results, dotted lines - predictions of  Eq.(\ref{shear}).}\label{fig:flow}
\end{figure}

To explore flow behavior near a hydrophobic surface with mobile charges in more details, we continue with varying the ratio $q_2/q_1$ from 0 to 3 at fixed $b/H=1.2$. The simulation results and theoretical predictions are given in Fig.~\ref{fig:flow} and are again in a good agreement (since the `flow' in the adsorbed layer is qualitatively the same as in Fig.~\ref{fig:test}(a) we do not show it here and below).
 We see that an apparent EO slip at the surface, $u_2$, increases with $q_2/q_1$, but a variety of physically different situations occurs depending on the value of this ratio. Uncharged hydrophobic surface induces an EO slip, and we see a manifestation of an outer `shear' flow. As discussed above, in the case of symmetric charges, $q_2/q_1=1$ we see no indication of a hydrodynamic slip.  Finally, for larger $q_2/q_1$ we again observe an outer `shear' flow.
These observations are well described by Eq.(\ref{shear}). It also suggests that if $q_2/q_1<1$ the hydrodynamic slippage amplifies  $u_2$ as compared to expected for a hydrophilic surface, where $u_2=u_1 q_2/q_1$, but when $q_2/q_1>1$ hydrodynamic slip inhibits the apparent EO slip. A key remark is that this amplification or inhibition is no longer dependent on the equilibrium properties of the EDL, but note that a rich outer EO behavior is accompanied by the unusual EDL dynamics. A charged hydrophobic surface actively participates in the flow-driving mechanism since it reacts electrostatically to the field by inducing a forward or backward inner EO flow.

\begin{figure}
\centering
  \includegraphics[scale=0.60]{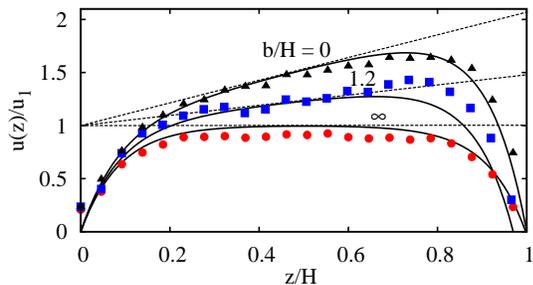}
  \caption{Fluid velocity profiles simulated at $\kappa H = 12$, $q_2/q_1= 2$, and $\mu=0$ (symbols). From top to bottom $b/H=0$, $1.2$, and $\infty$.  Solid curves show  theoretical results, dotted lines - predictions of Eq.(\ref{shear}). }\label{fig:Q}
\end{figure}

We now illustrate the influence of a hydrodynamic slip on EO flow in case $\mu=0$ (Fig.~\ref{fig:Q}). According to Eq.(\ref{zeta_general}) with the taken charge ratio, $q_2/q_1 = 2$, the apparent EO slip should be inhibited compared to a hydrophilic case, which is fully confirmed by our results. In case of $b/H=O(1)$ we observe a decrease in the outer `shear' EO flow and a corresponding apparent EO slip at the hydrophobic surface. However, in the limit of $b=\infty$ (a wetting film), we observe the `plug' outer flow (also reported before~\cite{choi.w:2010}), which reflects the EDL dynamics, where electrostatically active interface induces the strong inner flow opposite to the field.

\begin{figure}
\centering
 \includegraphics[scale=0.60]{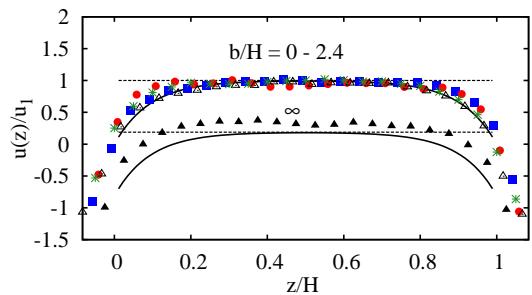}
  \caption{Fluid velocity profiles in a symmetric channel simulated at $\kappa H = 11$ and $\mu=0$ (symbols). Upper curves were simulated at $b/H=0$, $0.08$, $1.3$, $2.4$, bottom curve corresponds to $b/H=\infty$. Solid curves show theoretical results, upper dotted lines - predictions of Eq.(\ref{symmetric}) and lower - of Eq.(\ref{infinite}).
  }\label{fig:symmetric}
\end{figure}

Let us now turn to the EO properties of a symmetric channel with $\mu=0$ and plot in Fig.~\ref{fig:symmetric} the simulated EO velocities [related to $u_1$ expected in the no-slip case with $q_1=q_2$] for several values of the slip length. We see that outer EO flows simulated at several finite $b$ indeed coincide with the Smoluchowsky profile as predicted by Eq.(\ref{symmetric}), and are accompanied by the inner EO in the opposite direction. We have also explored what happens when $b=\infty$, and generally confirm a much smaller magnitude of a `plug' outer flow, also observed before~\cite{huang.dm:2008b}.

Finally, we can interpret a zeta-potential of a hydrophobic surface, which is naturally defined as  $\zeta_{2}=-u_{2} \eta / E_t \varepsilon$. In a thick asymmetric channel it is therefore $\zeta_2/\zeta_1=u_2/u_1$, described by  Eq.(\ref{zeta_general}).
However, if $b/H \ll 1$ Eq.(\ref{zeta_general}) reduces to

\begin{equation}\label{zeta_thick}
    \zeta_2 =   \frac{q_2 ( 1 + \mu \kappa b)}{\kappa\varepsilon}
\end{equation}
and zeta-potential becomes a characteristic of a hydrophobic surface solely. Eq.(\ref{zeta_thick}) is relevant for the understanding of highly debated zeta-potential measurements on free  interfaces of (not confined) bubbles and oil drops~\cite{creux.p:2009,marinova.kg:1996,Takahashi05}. Eq.(\ref{symmetric}) implies that a zeta-potential of a hydrophobic surface in a thick symmetric channel is also given by Eq.(\ref{zeta_thick}), except the case $b=\infty$ and $\mu=0$, where it becomes $\zeta_{2}=-2 u_{2} \eta / E_t \varepsilon \kappa H \simeq 0$.


 In conclusion, we have formulated an electro-hydrodynamic slip boundary condition and demonstrated that both confinement and mobility of surface charges has a dramatic effect on the dynamic properties of the EDL and EO flow.  Simple analytical formulae for the apparent EO slip and zeta-potential at the hydrophobic surface have been suggested, which resolve a number of paradoxes and confusions in the literature. Our results obtained for cases of immobile and fully mobile charges give rigorous upper and lower bounds on an EO slip for arbitrary hydrophobic surfaces given only the surface charge/potential and (any) slip lengths. These bounds constrain the attainable zeta-potential, and provide guidance for experimental measurements of $\mu$, which in some real systems could be confined in the interval from 0 to 1. Our study may be immediately extended to and/or for the challenging case of $\kappa H = O(1)$ and smaller~\cite{schoch.pb:2008,eijkel.jct:2005}, where the outer EO is absent.
Another fruitful direction could be to apply them to revisit calculations of an EO flow past superhydrophobic surfaces~\cite{belyaev.av:2011a,Squires08,bahga:2009}.

This research was partly supported by the Russian Foundation for Basic Research (grant 12-03-00916) and by the DFG through SFB 985.
The simulations were carried out using computational resources at the Moscow State University (`Lomonosov' and `Chebyshev').

\bibliographystyle{apsrev}
\bibliography{EO}

\end{document}